\documentclass[12pt]{amsart}
\usepackage{amsmath,amstext, amssymb,graphicx,fancybox,subfigure,amsthm}
\newtheorem{theorem}{Theorem}

\theoremstyle{definition}

\newtheorem{notation}[theorem]{Notation}

\theoremstyle{remark}

\def\id{\text{{\bf id}}}

\def\cG{\mathcal{G}}

\def\cA{\mathcal{A}}
\def\cF{\mathcal{F}}

\def\cB{\mathcal{B}}

\def\cD{\mathcal{D}}

\def\HH{\mathcal{H}}

\def\CC{{\Bbb C}}
\def\NN{{\Bbb N}}
\def\RR{{\Bbb R}}

\def\ff{\varphi}

\def\tr{\text{\rm{tr}}}

\def\tovert{\overset {\text{distr}}{\longrightarrow}}

\def\ee{\varepsilon}

\def\cH{\HH}

\title[Non-separable correlations]{On slow-fading non-separable correlation MIMO systems}
\author[R. Rashidi Far, T. Oraby, W. Bryc, R. Speicher]{ R. Rashidi Far \and T. Oraby \and W.
Bryc$^*$\and R. Speicher$^+$ }

\thanks{$^*$ Research partially supported by NSF grant
\#DMS-0504198.}
\thanks{$^+$ Research supported by a Discovery Grant and a Leadership
Support Initiative Award from the Natural Sciences and Engineering Research Council of
Canada and by a Killam Fellowship from the Canada Council for the Arts}
\thanks{R. Rashidi Far and R. Speicher are with the Department of
Mathematics and Statistics, Queen's University, Ontario, Canada K7L 3N6 reza,
speicher@mast.queensu.ca , T. Oraby and W. Bryc are with the Department of Mathematical
Sciences, University of Cincinnati, 28855, Campus Way PO Box 210025, Cincinnati, OH
45221-0025, USA, orabyt@math.uc.edu, wlodzimierz.bryc@uc.edu . }

\begin{document}

\setlength{\unitlength}{0.5cm}

\newsavebox{\NCzeZ}
\savebox{\NCzeZ}(2,2){
\thicklines
\put(0,0){\line(1,0){1}}
\put(0,0){\line(0,1){1}}
\put(1,0){\line(0,1){1}}
\put(0,1.7){\makebox(0,0){1}}
\put(1,1.7){\makebox(0,0){2}}}

\newsavebox{\NCvzZ}
\savebox{\NCvzZ}(3,3){
\thicklines
\put(0,0){\line(0,1){2}}
\put(0,0){\line(1,0){3}}
\put(3,0){\line(0,1){2}}
\put(1,1){\line(0,1){1}}
\put(1,1){\line(1,0){1}}
\put(2,1){\line(0,1){1}}
\put(0,2.7){\makebox(0,0){1}}
\put(1,2.7){\makebox(0,0){2}}
\put(2,2.7){\makebox(0,0){3}}
\put(3,2.7){\makebox(0,0){4}}}

\newsavebox{\NCveZ}
\savebox{\NCveZ}(3,3){
\thicklines
\put(0,0){\line(0,1){1}}
\put(0,0){\line(1,0){1}}
\put(1,0){\line(0,1){1}}
\put(2,0){\line(0,1){1}}
\put(2,0){\line(1,0){1}}
\put(3,0){\line(0,1){1}}
\put(0,1.7){\makebox(0,0){1}}
\put(1,1.7){\makebox(0,0){2}}
\put(2,1.7){\makebox(0,0){3}}
\put(3,1.7){\makebox(0,0){4}}}

\newsavebox{\NCsfZ}
\savebox{\NCsfZ}(5,4){ \thicklines \put(0,0){\line(0,1){3}} \put(0,0){\line(1,0){5}}
\put(5,0){\line(0,1){3}} \put(1,1){\line(0,1){2}} \put(1,1){\line(1,0){3}}
\put(4,1){\line(0,1){2}} \put(2,2){\line(1,0){1}} \put(2,2){\line(0,1){1}}
\put(3,2){\line(0,1){1}} \put(0.4,3.7){\makebox(0,0){$sb_1$}}
\put(1.4,3.7){\makebox(0,0){$sb_2$}} \put(2.4,3.7){\makebox(0,0){$sb_3$}}
\put(3.4,3.7){\makebox(0,0){$sb_4$}} \put(4.4,3.7){\makebox(0,0){$sb_5$}}
\put(5,3.7){\makebox(0,0){$s$}}}

\newsavebox{\NCsvZ}
\savebox{\NCsvZ}(5,3){ \thicklines \put(0,0){\line(0,1){2}} \put(0,0){\line(1,0){5}}
\put(5,0){\line(0,1){2}} \put(1,1){\line(0,1){1}} \put(1,1){\line(1,0){1}}
\put(2,1){\line(0,1){1}} \put(3,1){\line(1,0){1}} \put(3,1){\line(0,1){1}}
\put(4,1){\line(0,1){1}} \put(0.4,2.7){\makebox(0,0){$sb_1$}}
\put(1.4,2.7){\makebox(0,0){$sb_2$}} \put(2.4,2.7){\makebox(0,0){$sb_3$}}
\put(3.4,2.7){\makebox(0,0){$sb_4$}} \put(4.4,2.7){\makebox(0,0){$sb_5$}}
\put(5,2.7){\makebox(0,0){$s$}}}

\newsavebox{\NCsdZ}
\savebox{\NCsdZ}(5,3){ \thicklines \put(0,0){\line(0,1){2}} \put(0,0){\line(1,0){3}}
\put(3,0){\line(0,1){2}} \put(1,1){\line(0,1){1}} \put(1,1){\line(1,0){1}}
\put(2,1){\line(0,1){1}} \put(4,0){\line(0,1){2}} \put(4,0){\line(1,0){1}}
\put(5,0){\line(0,1){2}} \put(0.4,2.7){\makebox(0,0){$sb_1$}}
\put(1.4,2.7){\makebox(0,0){$sb_2$}} \put(2.4,2.7){\makebox(0,0){$sb_3$}}
\put(3.4,2.7){\makebox(0,0){$sb_4$}} \put(4.4,2.7){\makebox(0,0){$sb_5$}}
\put(5,2.7){\makebox(0,0){$s$}}}

\newsavebox{\NCszZ}
\savebox{\NCszZ}(5,4){ \thicklines \put(0,0){\line(0,1){2}} \put(0,0){\line(1,0){1}}
\put(1,0){\line(0,1){2}} \put(2,0){\line(0,1){2}} \put(2,0){\line(1,0){3}}
\put(5,0){\line(0,1){2}} \put(3,1){\line(0,1){1}} \put(3,1){\line(1,0){1}}
\put(4,1){\line(0,1){1}} \put(0.4,2.7){\makebox(0,0){$sb_1$}}
\put(1.4,2.7){\makebox(0,0){$sb_2$}} \put(2.4,2.7){\makebox(0,0){$sb_3$}}
\put(3.4,2.7){\makebox(0,0){$sb_4$}} \put(4.4,2.7){\makebox(0,0){$sb_5$}}
\put(5,2.7){\makebox(0,0){$s$}}}

\newsavebox{\NCseZ}
\savebox{\NCseZ}(5,4){ \thicklines \put(0,0){\line(0,1){1}} \put(0,0){\line(1,0){1}}
\put(1,0){\line(0,1){1}} \put(2,0){\line(0,1){1}} \put(2,0){\line(1,0){1}}
\put(3,0){\line(0,1){1}} \put(4,0){\line(0,1){1}} \put(4,0){\line(1,0){1}}
\put(5,0){\line(0,1){1}} \put(0.4,1.7){\makebox(0,0){$sb_1$}}
\put(1.4,1.7){\makebox(0,0){$sb_2$}} \put(2.4,1.7){\makebox(0,0){$sb_3$}}
\put(3.4,1.7){\makebox(0,0){$sb_4$}} \put(4.4,1.7){\makebox(0,0){$sb_5$}}
\put(5,1.7){\makebox(0,0){$s$}}}

\date{}

\begin{abstract}
In a frequency selective slow-fading channel in a MIMO system, the channel matrix is of
the form of a block matrix. We propose a method to calculate the limit of the eigenvalue
distribution of block matrices if the size of the blocks tends to infinity. We will also
calculate the asymptotic eigenvalue distribution of $HH^*$, where the entries of $H$ are
jointly Gaussian, with a correlation of the form $E[h_{pj}\bar h_{qk}]= \sum_{s=1}^t
\Psi^{(s)}_{jk}\hat\Psi^{(s)}_{pq}$ (where $t$ is fixed and does not increase with the
size of the matrix). We will use an operator-valued free probability approach to achieve
this goal. Using this method, we derive a system of equations, which can be solved
numerically  to compute the desired eigenvalue distribution.
\end{abstract}

\maketitle


\section{Introduction}
With the introduction of some sophisticated communication techniques such as CDMA
(Code-Division Multiple-Access) and MIMO (Multiple-Input Mul\-tiple-Output), the
communications  community has been looking into analyzing different aspects of these
systems, ranging from the channel capacity to the structure of the receiver. It has been
shown that the channel matrix plays a key role in the capacity of the channel
\cite{Foschini-Gans-98,Verdu-86} as well as in the structure of the optimum receiver
\cite{Madhow-Honig-94,Verdu-98}. More precisely, the eigenvalue distribution of the
channel matrix is the factor of interest in different applications.

Free probability \cite{VDN-92,Hiai-Petz,Nica-Speicher} and random matrix theory have proven to
provide the right kind of tools in tackling such kind of problems
\cite{Tulino-Verdu,Edelman-Rao-06}. For example, Tse and Zeitouni \cite{Tse-Zeitouni-00}
applied random matrix theory to study linear multiuser receivers, Moustakas  {\em et. al.}
\cite{Moustakas-Simon-Sengupta-03} applied it to calculate the capacity of a MIMO channel.
M\"uller
{Muller-02,Muller-02a} employed it in calculating the eigenvalue distribution of
a particular fading channel and later Debbah and  M\"uller \cite{Debbah-Muller-05} applied it
in MIMO channel modeling.

\begin{figure}[htb]
\centering{
  \includegraphics[width=12cm]{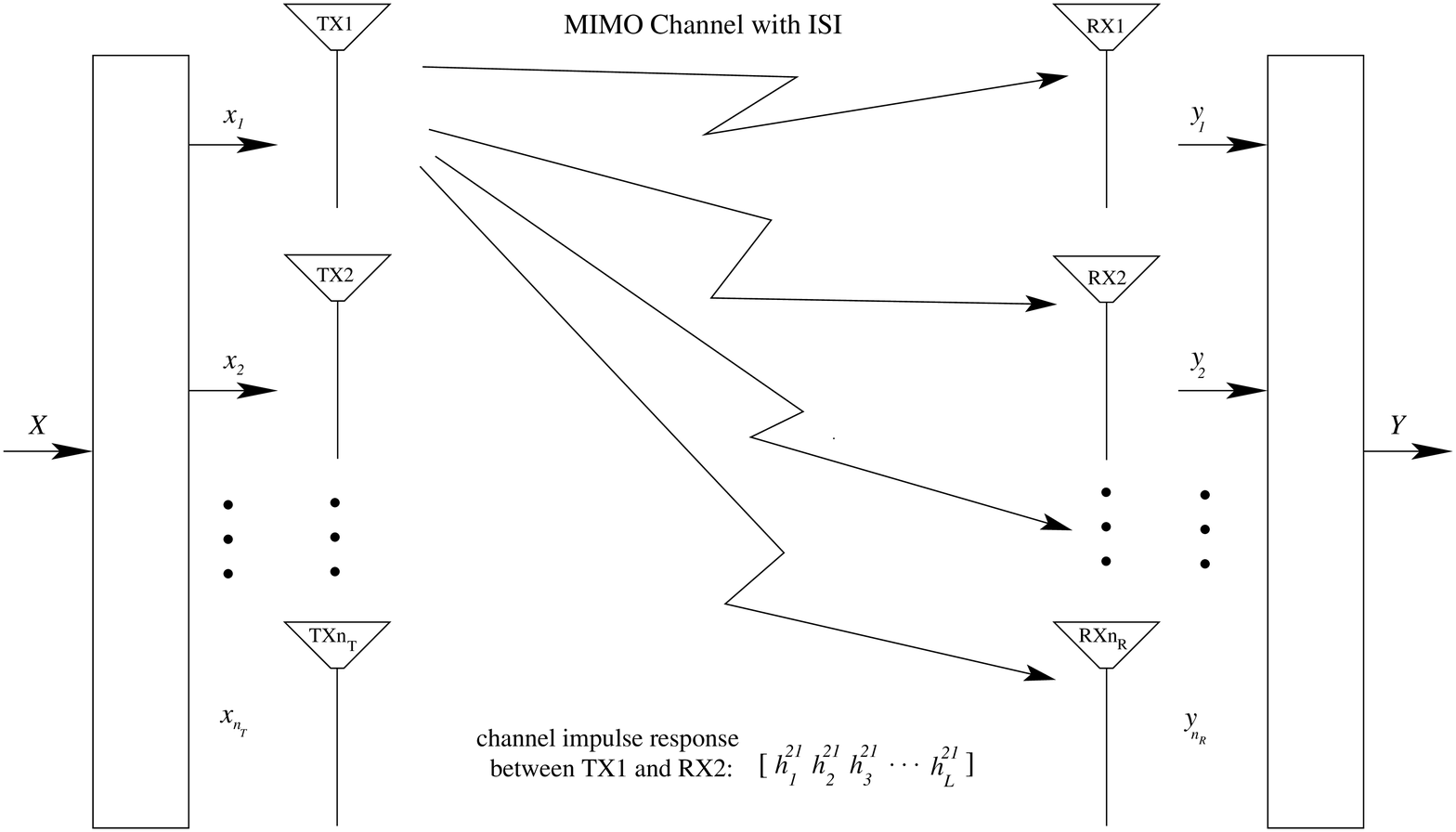}}
\caption{\label{fig:MIMOblck}Block diagram of a MIMO system with ISI.}
\end{figure}

There are, however, also many interesting (more realistic) models for the channel matrix,
which are not directly accessible with the usual free probability or random matrix
techniques. Let us be a bit more specific on such examples. For a MIMO wireless system
with $n_T$ transmitter antenna and $n_R$ receiver antenna, the received signal at time
index $n$, $Y_n=\left[y_{1,n},\cdots,y_{n_R,n}\right]^T$, will be as follows:
\begin{eqnarray}
  Y_n=HX_n+N_n,
\end{eqnarray}
where $H$ is the channel matrix, $X_n=\left[x_{1,n},\cdots,x_{n_T,n}\right]^T$ is  the
transmitted signal at time $n$ and $N_n$ is the noise signal. The channel matrix entries
$h_{ij}$ reflect the channel effect on the signal transmitted  from antenna $j$ in the
transmitter and received at antenna $i$ in the receiver. In a more realistic channel
modeling, one may consider the Intersymbol-Interference (ISI) \cite[Chapter
2]{Larsson-Stoica-03}. In this case, the channel impulse response between the transmitter
antenna $j$ and the receiver antenna $i$ is a  vector
$h_{ij}=\left[\begin{array}{ccccc}h^{(ij)}_1& h^{(ij)}_2&
\cdots&h^{(ij)}_{L-1}&h^{(ij)}_{L}\end{array}\right]^T$ where $L$ is the length of the
impulse response of the channel (number of the taps). Consequently, the channel matrix
for a signal frame of $K$ will be as follows:
\begin{eqnarray}\label{eq:channel}
  H&=&\left[\
    \begin{array}{ccccccccc}
      A_1&A_2&\cdots&A_{L}&{\bf 0}&{\bf 0}&&\cdots&{\bf 0}\\
      {\bf 0}&A_1&A_2&\cdots&A_{L}&{\bf 0}&&\cdots&{\bf 0}\\
      {\bf 0}&{\bf 0}&A_1&A_2&\cdots&A_{L}&{\bf 0}&\cdots&\vdots\\
      \vdots&\vdots&\ddots&\ddots&\cdots&&\ddots&\cdots&{\bf 0}\\
      {\bf 0}&{\bf 0}&\cdots&\cdots&{\bf 0}&A_1&A_2&\cdots&A_{L}
    \end{array}
    \right],
\end{eqnarray}
where there are $K-1$ zero-matrices in each row and
$A_l=(h^{\left(ij\right)}_l)_{{\scriptstyle i=1,\cdots,n_R}\atop{\scriptstyle
j=1,\cdots,n_T}}$ (see Fig.~\ref{fig:MIMOblck} for the block diagram). To calculate the
capacity of such a channel, one needs to know the eigenvalue distribution of the $HH^*$
\cite{Tulino-Verdu}.

The above random matrix falls into the class of random matrices where one has
correlations between the entries of $H$. Whereas random matrices with independent entries
are quite well understood and there exist many analytic results on their asymptotic
eigenvalue distribution, only very special cases of the correlated situation could be
treated in the literature. The most prominent of those is the case of separable
correlation, where the covariance between the entries of $H=(h_{ij})_{i,j}$ factorizes as
$E[h_{pj}\bar h_{qk}]= \Psi_{jk}^T\Psi_{pq}^R$, where $\Psi^T$ and $\Psi^R$ are matrices
describing the transmit and the receive correlation, respectively. Our block matrix $H$
from Eq. \eqref{eq:channel} does not fall into this class.

In this paper, we will show how a more general version of free probability theory, so-called
``operator valued free probability theory'' allows to deal with more general situations of
correlated entries. In particular, we will treat the case of block matrices, as the above $H$
from Eq. \eqref{eq:channel}, and also extend results from \cite{CTKV} from the case of
separable correlations to the more general situation
\begin{equation}
E[h_{pj}\bar h_{qk}]= \sum_{s=1}^t \Psi^{(s)}_{jk}\hat\Psi^{(s)}_{pq}.\end{equation} The
results for the asymptotic eigenvalue distribution of $HH^*$ for these two cases are
stated in Section \ref{sec:statements}, in Theorems \ref{thm:H-block} and
\ref{thm:H-non-sep}. The proof of Theorem \ref{thm:H-block} is given in Section
\ref{sec:three}, as a consequence of a corresponding statement, Theorem \ref{thm:square},
for selfadjoint matrices $X$. We will show that these selfadjoint matrices are
asymptotically described by operator-valued semicircular elements, and the equation
describing the limiting Cauchy transform of $X$ follows then from the general theory of
semicircular elements. In Appendix I we state the main notions and results in relation
with operator-valued semicircular elements. In Appendix II we prove Theorem
\ref{thm:H-non-sep}, again by showing that a corresponding selfadjoint matrix $X$ is
asymptotically an operator-valued semicircular element. In Appendix III we state a more
general version of Theorem \ref{thm:square}, for the situation where the blocks are not
necessarily square matrices.

\section{Asymptotic eigenvalue distribution of $HH^*$}\label{sec:statements}

In this section we will present our main results on the asymptotic eigenvalue
distribution for $HH^*$ where $H$ is a non-selfadjoint Gaussian random matrix with some
specific kind of correlation between its entries. We will treat the block matrix case and
the case of non-separable correlation. The proof of these theorems will be provided in
the next section and in the appendix. An application to the asymptotic eigenvalue
distribution of channel matrices of the form \eqref{eq:channel} will be given in Section
\ref{sec:results}.

\subsection{$HH^*$ for block matrices}

Our main theorem on block channel matrices is the following. For used notation, see
Section \ref{sec:notations}.

\newcommand{\rrr}{b}
\newcommand{\sss}{a}

\begin{theorem}\label{thm:H-block}
Fix natural $\rrr $ and $\sss $ and a
real valued ``covariance function'' $\tau(i,k;j,l)$ such that $\tau(i,k;j,l)=\tau(j,l;i,k)$, $i,j=1,\dots,\sss ;
k,l=1,\dots,\rrr $. Assume, for $N\in\NN$, that
$\{h_{rp}^{(i,k)}\mid i=1,\dots,\sss ,\,k=1,\dots,\rrr ,\,r,p=1,\dots,N\}$ are
jointly Gaussian
complex random variables, with the prescription of mean zero
 and covariance
\begin{equation}\label{eq:cov-square}
E[h_{rp}^{(i,k)} \bar h_{sq}^{(j,l)}]=\frac 1{(\rrr +\sss )N} \delta_{rs}\delta_{pq}\cdot
\tau(i,k;j,l).
\end{equation}
We also assume circular complex Gaussian  law, i.e, $E[(h_{rp}^{(i,k)} )^2]=0$.

Consider now block matrices $H_N=(H^{(i,k)})_{i=1,\dots,\sss \atop k=1,\dots,\rrr }$ , where, for each
$i=1,\dots,\sss $ and $k=1,\dots,\rrr $, the blocks are given by
$H^{(i,k)}=\bigl(h^{(i,k)}_{rp}\bigr)_{r,p=1}^N$.

Then, for $N\to\infty$, the $\sss N\times \sss N$ matrix $H_NH_N^*$ has almost surely a
limiting eigenvalue distribution whose Cauchy transform $G(z)$ is determined by
$$G(z)=\tr_\sss (\cG_1(z)),$$
where $\cG_1(z)$ is an $M_\sss (\CC)$-valued analytic function on the upper complex half plane,
which is uniquely determined by the facts that
\begin{equation}
\lim_{|z|\to\infty, \Im(z)>0}z\cG_1(z)=I_\sss , \end{equation}
 and that it satisfies
for all $z$ in the upper complex half plane the matrix equation
\begin{equation}\label{eq:poisson}
z\cG_1(z)= I_\sss + \eta_1\Bigl(\bigl( I_\rrr -\eta_2\bigl(\cG_1(z)\bigr)\bigr)^{-1}
\Bigr)\cdot \cG_1(z),
\end{equation}
where
$$\eta_1:M_\rrr (\CC)\to M_\sss (\CC) \qquad\text{and}\qquad
\eta_2:M_\sss (\CC)\to M_\rrr (\CC)$$ are the covariance mappings given by
\begin{equation}
  \label{eq:eta1}
  \left[\eta_1(D)\right]_{ij}:=\frac 1{\rrr +\sss }\sum_{k,l=1}^\rrr  \tau(i,k;j,l)\cdot [D]_{kl}
\end{equation}
and
\begin{equation}
    \label{eq:eta2}
\left[\eta_2(D)\right]_{kl}:=\frac 1{\rrr +\sss }\sum_{i,j=1}^\sss  \tau(i,k;j,l)\cdot [D]_{ji}.
\end{equation}

\end{theorem}

The proof of this theorem will be given in Section \ref{sec:three}, by reducing it to
Theorem
\ref{thm:square}.

In Section \ref{sec:results} we will use this to analyze the asymptotic eigenvalue
distribution of $HH^*$ for the channel matrix from Eq. \eqref{eq:channel}.

\subsection{$HH^*$ for non-separable correlated fading}

In \cite{CTKV}, MIMO wireless systems under correlated fading were analyzed by asymptotic
analysis of the eigenvalue distribution of $H_nH_n^*$, where the entries of the $n\times
n$ random matrix $H=1/\sqrt{n}(h_{ij})_{i,j=1}^n$ were assumed as jointly Gaussian with
the following covariance structure:
$$E[h_{pj}\bar h_{qk}]= \Psi_{jk}^T\Psi_{pq}^R,$$
where $\Psi^T$ and $\Psi^R$ are Hermitian positive-definite matrices describing the
transmit and the receive correlation, respectively. The assumption on $\Psi^T$ and
$\Psi^R$ is that both have a limiting eigenvalue distribution.

We will now show how operator-valued free probability theory can be used to analyze a
generalization of this to the case
\begin{equation}\label{eq:psi}
E[h_{pj}\bar h_{qk}]=\sum_{s=1}^t \Psi^{(s)}_{jk}\hat\Psi^{(s)}_{pq}.
\end{equation}
The number $t$ of summands is here fixed and does not depend on $n$. As before, one needs
the existence of the limiting joint distribution of the $\Psi$'s and the limiting joint
distribution of the $\hat\Psi$'s. Mixed moments in $\Psi$ and $\hat\Psi$ do not play a
role for the result on $HH^*$.

This situation is treated in the next theorem, which we will prove in Appendix II,
Section \ref{proof:6-2}. Some of the basic notions from free probability which are used
in the formulation of the theorem are defined in Appendix I. As in \cite{CTKV} we will
restrict here, for notational simplicity, to the case of a square $H$. By invoking ideas
from \cite{Benaych-Georges-05}, one can also extend the results to rectangular $H$.

\begin{theorem}\label{thm:H-non-sep}
Assume that $h_{ij}$ ($i,j\in\NN$) are jointly Gaussian complex random variables
with mean zero and covariance given by
\eqref{eq:psi} for some $t\geq 1$ and some positive-definite matrices $\Psi_{ij}^{(s)}$
and $\hat \Psi_{ij}^{(s)}$ ($s=1,\dots,t$). We also assume circular complex Gaussian
law, i.e, $E[(h_{rp} )^2]=0$. We assume that,
as $n\to\infty$, the
$\bigl((\Psi^{(s)}_{i,j}),(\hat\Psi^{(s)}_{i,j})\bigr)_{s=1,\dots,t}$ converge in
distribution  to some elements $(\Psi_s,\hat\Psi_s)_{s=1,\dots,t}$ in some
non-commutative probability space $(\cB,\ff)$.

We denote by $\cB_1\subset\cB$ the algebra generated by $\Psi_1$, \dots, $\Psi_t$ and by
$\cB_2\subset\cB$ the algebra generated by $\hat\Psi_1$, \dots, $\hat\Psi_t$. Furthermore
we define
\begin{equation}
\eta_1:\cB_1\to\cB_2,\qquad \eta_1(b):=\sum_{s=1}^t \hat\Psi_s \ff(b\Psi_s) \end{equation}
and
\begin{equation}
\eta_2:\cB_2\to\cB_1,\qquad \eta_2(b):=\sum_{s=1}^t \Psi_s \ff(b\hat\Psi_s).
\end{equation}
We consider now
$$H_n:=\frac 1{\sqrt {n}} \bigl(h_{ij}\bigr)_{i,j=1}^n$$
Then the eigenvalue distribution of $H_nH_n^*$ converges almost surely to a limiting
distribution whose Cauchy transform $G$ is given by $G(z)=\ff(\cG_1(z))$, where $\cG_1$
is the solution of the equation
\begin{equation}\label{eq:fading-solution}
z\cG_1(z)= \id+ \sum_{s_1=1}^t\Psi_{s_1}\ff\left(\Bigl( \id-\sum_{s_2=1}^t \hat\Psi_{s_2}
\ff\bigl(\cG_1(z)\Psi_{s_2}\bigr) \Bigr)^{-1}\hat\Psi_{s_1}\right)\cdot \cG_1(z).
\end{equation}
\end{theorem}

One should note that the solution $\cG_1(z)$ of the above fixed point equation lies in
the algebra $\cB_1$ and that its value does not depend on mixed moments between the
$\Psi_s$'s and the $\hat\Psi_s$'s. By results from \cite{HFS} (as outlined in Appendix I,
Section \ref{sec:semi}), there exists, for each $z\in\CC^+$ a unique solution of equation
\eqref{eq:fading-solution} with the right positivity property.

Note also that Eq. \eqref{eq:fading-solution} reduces in the case $t=1$ to the fixed
point equation in Theorem IV.2 in \cite{CTKV}.

\section{Asymptotic eigenvalue distribution for selfadjoint block matrices}\label{sec:three}

Our Theorem \ref{thm:H-block} on the asymptotic eigenvalue distribution of $HH^*$ for a
block matrix $H$ follows from a corresponding statement for a selfadjoint block matrix
$X$, which also has Gaussian entries with correlations. The reduction to the selfadjoint
case can be achieved by the well-known trick of going over to
\begin{equation}\label{eq:X}X=\begin{bmatrix}
0& H\\
H^*&0
\end{bmatrix}.\end{equation}

In this section we will state the selfadjoint version of Theorem \ref{thm:H-block} and
show how it implies the result for $HH^*$.

\subsection{Selfadjoint block matrices}
Let us consider the selfadjoint version of Theorem \ref{thm:H-block}. We will here
restrict to the situation where all blocks are square matrices of the same size. For some
applications it might actually be better to allow also blocks of a rectangular size
(which, of course, have to fit together to form a big square matrix). There is a
straightforward generalization of the following theorem to that situation; we will state
this in Appendix III.

\begin{theorem}\label{thm:square}
Fix a natural $d$ and a  ``covariance function'' $\sigma$ which satisfies
\begin{equation}
  \label{symmetry}
  \sigma(i,j;k,l)=\overline{\sigma(k,l;i,j)}\qquad
\end{equation}
for all $i,j,k,l=1,\dots,d$. Assume, for $N\in\NN$, that $\{a_{rp}^{(i,j)}\mid i,j=1,\dots,d,\,
r,p=1,\dots,N\}$ are jointly Gaussian random variables, with
$$a^{(i,j)}_{rp}=\overline{a_{pr}^{(j,i)}}\qquad\text{for all $i,j=1,\dots,d$,
$r,p=1,\dots,N$}$$ and the prescription of mean zero and covariance
\begin{equation}
E[a_{rp}^{(i,j)} a_{qs}^{(k,l)}]=\frac 1{dN} \delta_{rs}\delta_{pq}\cdot \sigma(i,j;k,l).
\end{equation}
Consider now block matrices $X_N=(A^{(i,j)})_{i,j=1}^d$ , where, for each $i,j=1,\dots,d$, the
blocks are given by $A^{(i,j)}=\bigl(a^{(i,j)}_{rp}\bigr)_{r,p=1}^N$.

Then, for $N\to\infty$, the $dN\times dN$ matrix $X_N$ has almost surely a limiting
eigenvalue distribution whose Cauchy transform $G(z)$ is determined by
\begin{equation}
G(z)=\tr_d(\cG(z)), \end{equation}
where $\cG(z)$ is an $M_d(\CC)$-valued analytic function on
the upper complex half plane, which is uniquely determined by the facts that
\begin{equation}\label{eq:cG lim}
\lim_{|z|\to\infty, \Im(z)>0}z\cG(z)=I_d, \end{equation}
 and that it satisfies
for all $z$ in the upper complex half plane the matrix equation
\begin{equation}\label{eq:square}
 z \cG(z)= I_d + \eta(\cG(z))\cdot \cG(z),
\end{equation}
where $\eta:M_d(\CC)\to M_d(\CC)$ is the covariance mapping
\begin{equation}
\left[\eta(D)_{i,j=1}^d)\right]_{ij}:=\frac 1d\sum_{k,l=1}^d \sigma(i,k;l,j)\cdot [D]_{kl}.
\end{equation}
\end{theorem}

The proof of Theorem \ref{thm:square} is given in Appendix II. Let us here just point out
that the determining equation \eqref{eq:square} is actually the equation for an
operator-valued semicircular element; thus, one essentially has to realize that $X_N$
converges to a suitably chosen operator-valued semicircular element.

Theorem \ref{thm:square} has also some interest of its own; for an application to some
selfadjoint block matrix problems from \cite{Tamer} see \cite{RTBS}. Here we will just
use it to prove our Theorem \ref{thm:H-block}

\subsection{Proof of Theorem \ref{thm:H-block}}\label{proof:H-block}
Let us consider
matrices $H_N$ as in Theorem \ref{thm:H-block}. For clarity of notation, we will in the
following suppress the index $N$. The calculation of the eigenvalue distribution of
$HH^*$ can be reduced to the situation treated in the previous section by the following
trick. Consider
$$
X=\begin{bmatrix}
0& H\\
H^*&0
\end{bmatrix}.$$
With $d=\rrr +\sss $, this is a selfadjoint $dN\times dN$-matrix and can be viewed as a $d\times
d$-block matrix of the form considered in Theorem \ref{thm:square}; thus we can use this
to get the asymptotic eigenvalue distribution of $X$.

The only remaining question is how to relate the eigenvalues of $X$ with those of $HH^*$.
This is actually quite simple, we only have to note that all the odd moments of $X$ are
zero and
$$X^2=\begin{bmatrix}
HH^*&0\\
0& H^*H
\end{bmatrix}$$
Thus the eigenvalues of $X^2$ are the eigenvalues of $HH^*$ together with the eigenvalues
of $H^*H$. (One might also note $HH^*$ is an $\sss N\times \sss N$ and $H^*H$ is an $\rrr N\times \rrr N$
matrix. Assuming that $\sss <\rrr $ (otherwise exchange the role of $H$ and $H^*$) we have then
that the eigenvalues of $H^*H$ are the eigenvalues of $HH^*$ plus $(\rrr -\sss )N$ additional
zeros. However, we will not need this information in the following.)

So we should rewrite our equation for the Cauchy transform $G_X$ of $X$ in terms of the
Cauchy transform $G_{X^2}$ of $X^2$. Since $X$ is even, both are related by
$$z\cdot G_{X^2}(z^2)=G_X(z).$$
By noting that the operator-valued Cauchy transform $\cG(z)$ of $X$ depends, up to an
overall factor $1/z$, only on $z^2$, we can introduce a quantity $\cH$ by
$$z\cdot\cH(z^2)=\cG(z).$$
Then with $n=dN$ we have
\begin{equation}\label{eq:eins}
\lim_{n\to\infty} G_{X^2}(z)=\tr_d[\cH(z)],
\end{equation}
and the equation (\ref{eq:square}) for $\cG$ becomes
\begin{equation}\label{eq:zwei}
z \cH(z)= I_d+z\eta\bigl(\cH(z)\bigr)\cdot \cH(z).
\end{equation}
It is fairly easy to see that
 the covariance mapping $\eta:M_{\rrr +\sss }(\CC)\to
M_{\rrr +\sss }(\CC)$ of $X$ splits according to
$$\eta:\begin{bmatrix}
D_1&D_3\\
D_4&D_2
\end{bmatrix}\mapsto
\begin{bmatrix}
\eta_1(D_2)&0\\
0&\eta_2(D_1)
\end{bmatrix},
$$
where
$$\eta_1:M_\rrr (\CC)\to M_\sss (\CC) \qquad\text{and}\qquad
\eta_2:M_\sss (\CC)\to M_\rrr (\CC)$$ are the two covariance mappings for $H$ as in Theorem
\ref{thm:H-block}.
Therefore, our  $(\rrr +\sss )\times (\rrr +\sss )$ matrix $\cH$ decomposes as a
$2\times 2$-block matrix
$$\cH(z)=\begin{bmatrix}
\cG_1(z)& 0\\
0& \cG_2(z)
\end{bmatrix}$$
where $\cG_1$ and $\cG_2$ are $M_\sss (\CC)$-valued and $M_\rrr (\CC)$-valued,
respectively, analytic functions in the upper complex half plane. Then one has
\begin{equation}\label{eq:vier}
\lim_{N\to\infty}G_{HH^*}(z)=\tr_\sss \bigl(\cG_1(z)\bigr).
\end{equation}
and
\begin{equation}\label{eq:viera}
\lim_{N\to\infty}G_{H^*H}(z)=\tr_\rrr \bigl(\cG_2(z)\bigr).
\end{equation}

The equation (\ref{eq:zwei}) for $\cH$ splits now into the two equations
$$z\cG_1(z)=
I_\sss +z\eta_1\bigl(\cG_2(z)\bigr)\cdot \cG_1(z)$$ and
$$z\cG_2(z)=
I_\rrr +z\eta_2\bigl(\cG_1(z)\bigr)\cdot \cG_2(z).$$ One can eliminate $\cG_2$ from those
equations by solving the second equation for $\cG_2$ and inserting this into the first
equation, yielding Eq. \eqref{eq:poisson}.

\section{Results and Discussion}\label{sec:results}
Our theorems give us the Cauchy transform $G$ of the asymptotic eigenvalue distribution $HH^*$
of the considered matrices in the form $G(z)=\tr_r(\cG_1(z))$, where $\cG_1(z)$ is a solution
to the matrix equation \eqref{eq:poisson} or \eqref{eq:fading-solution}. Usually, it is more
convenient to deal with the equation \eqref{eq:zwei} for the corresponding selfadjoint matrix
$X$.

We recover the corresponding eigenvalue distribution $\mu$ from $G$ in the usual way, by
invoking Stieltjes inversion formula
\begin{equation}
  \label{eq:Stielties Inversion}
  d\mu(x)=-\frac 1\pi \lim_{\ee\searrow 0}\Im G(x+i\ee)dx,
\end{equation} where
the limit is weak convergence of measures.

Usually, there is no explicit solution for our matrix equations, so that we have to rely on
numerical methods for solving those. Note that we do not get directly an equation for $G$. We
first have to solve the matrix equation, then take the trace of this solution. Thus, in terms
of the entries of our matrix $\cG_1$ or $\cH$, we face a system of quadratic equations which we
solve numerically, either by using Newton's algorithm \cite{NumericalRecipes-92} or by
iterations as in Eq.~\eqref{eq:iteration}.

\subsection{Example: ISI channel matrix}
In this section we want to specify our general theorems to the case of the ISI channel matrices
as appearing in Eq.~\eqref{eq:channel}. For simplicity we treat the case of square blocks.

\subsubsection{Proposition}\label{prop:wireless}
Let $H_N$ be the channel matrix from Eq.~\eqref{eq:channel} with $n_R=n_T=:N$, such that
each entry $h_l^{ij}$ has variance 1. Put $d=2K+L-1$, $n=Nd$. As $N\to\infty$ the
spectral law of $H_NH_N^*/n$ converges with probability one to a deterministic
probability  measure which is a mixture of $K$ densities with Cauchy transform
\begin{equation}
\lim_{N\to\infty}G_{HH^*/n}(z)=\frac{1}{K}\sum_{j=1}^K f_j(z). \end{equation}
Functions $f_j$
are each a Cauchy transform of a probability measure and the following conditions hold.
\begin{enumerate}
\item $f_j=f_{K+1-j} \mbox{ for } 1\leq j\leq K $
\item The diagonal matrix $\cG_1=\mbox{diag}(f_1,\dots,f_K)$ satisfies equation \eqref{eq:poisson}
with $\eta_1:M_{K+L-1}(\CC)\to M_K(\CC)$ given by
\begin{equation}
  \label{eq:MIMO_eta1}
  [\eta_1(D)]_{ij}=\frac{1}{L+2K-1}\sum_{k=1}^K[D]_{i+k,j+k},\, 1\leq i,j\leq K
\end{equation}
and with $\eta_2: M_K(\CC) \to M_{K+L-1}(\CC)$ such that on the diagonal we have
\begin{equation}
  \label{eq:MIMO_eta2}
  [\eta_2(D)]_{jj}=\frac{1}{L+2K-1}\sum_{k=\max\{1,j-L+1\}}^{\min\{j,K\}}[D]_{j+k,j+k},\, 1\leq j\leq K+L-1.
\end{equation}
\end{enumerate}

\begin{proof}
We note that the only non-zero values of $\tau$ are
$\tau(i,j;k,j+k-i)=1$ when $1\leq i,k\leq K$, $i\leq j\leq i+L-1$. Therefore \eqref{eq:eta1} gives
\eqref{eq:MIMO_eta1} and \eqref{eq:eta2} gives
\eqref{eq:MIMO_eta2}.

 From  \eqref{eq:MIMO_eta1} and \eqref{eq:MIMO_eta2} we see that $\eta$ maps diagonal matrices into
diagonal matrices, so the solution $\cH$ of equation \eqref{eq:zwei} must be diagonal,
 $$\cH(z)=\mbox{diag}(f_1,f_2,\dots,f_K,g_1,g_2,\dots,g_{K+L-1}).$$
We now note that the symmetry conditions $f_j=f_{K+1-j}$ and $g_j=g_{K+L-j}$ are preserved
under the mapping $D\mapsto I_d+\eta(D)\cdot D$, therefore the same symmetries must be satisfied by the solution
$\cH$.
Thus $\cG_1=\mbox{diag}(f_1,\dots,f_K)$
satisfies \eqref{eq:poisson} and $f_j=f_{K+1-j}$ as claimed.
\end{proof}
\subsubsection{Example}
As a concrete example we consider a MIMO system with ISI ($L=4$) and frame size of 4 ($K=4$):
\begin{equation}
H_N=\left[\begin{array}{*{20}c}
   A & B & C & D & 0 & 0 & 0 \\ 0 & A & B & C &
   D & 0 & 0 \\ 0 & 0 & A & B & C & D & 0 \\ 0 &
   0 & 0 & A & B & C & D \\
   \end{array}\right],
\end{equation}
where $A,B,C,D$ are independent non-selfadjoint Gaussian $N\times N$-random
matrices. It is also assumed that the impulse response of the channel from any transmitter
antenna to any receiver antenna is identical and equal to $\left[\begin{array}{cccc}1& 1& 1&
1\end{array}\right]$. In this case $K=L=4$,
$$\cG_1(z)=\mbox{diag}(f_1(z),f_2(z),f_2(z),f_1(z)),$$
$$[\eta_1(D)]_{ii}=\frac{1}{11}\sum_{j=i}^{i+3}[D]_{jj}\; \mbox{ for } i=1,2,$$
$$
\eta_2(\cG_1)=\frac{1}{11}\mbox{diag}(f_1,f_1+f_2, f_1+2f_2,2f_1+2f_2)
$$
and \eqref{eq:poisson} yields the following system of equations.
\begin{eqnarray*}
z &=& \frac1{f_1} + \frac{1}{11 - f_1}+\frac{1}{11 - f_1 - f_2}
+\frac{1}{11 - f_1 - 2f_2} + \frac{1}{11 - 2f_1 - 2f_2},\\
 z &=&
\frac1{f_2} + \frac1{11 - f_1 - f_2} +\frac2{11 - f_1 - 2f_2} +
\frac1{11 - 2f_1 - 2f_2}.
\end{eqnarray*}

\begin{figure}[hbt]\centering{
\includegraphics[width=8cm]{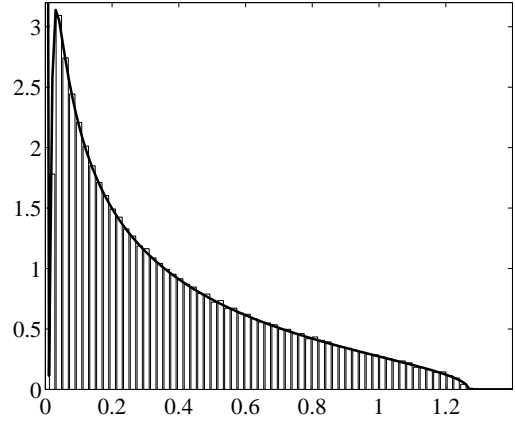}}
\caption{\label{ABCD000}Superimposed theoretical density of the eigenvalues of complex
normal $H_nH_n^*/n$ for a channel with ISI $L=4$ and a MIMO system $n_R=n_T$ with frame
length of $K=4$ over its histogram for $N=100$, based on $100$ realizations.}
\end{figure}

The limiting Cauchy transform is $G_{HH^*}(z)=(f_1+f_2)/2$.  We use Newton's algorithm to
solve this quadratic system of equations; the match between this solution and simulations
is shown in Fig.~\ref{ABCD000}.

\subsection{Convergence speed of capacity}
The results developed in this manuscript are good assets to study the asymptotic
behaviour of slow-fading non-separable correlation MIMO channels when $N\rightarrow
\infty$ but the authenticity of these results for limited $N$ is also of interest in
practice.

\begin{figure}[htb]
\includegraphics[width=12cm]{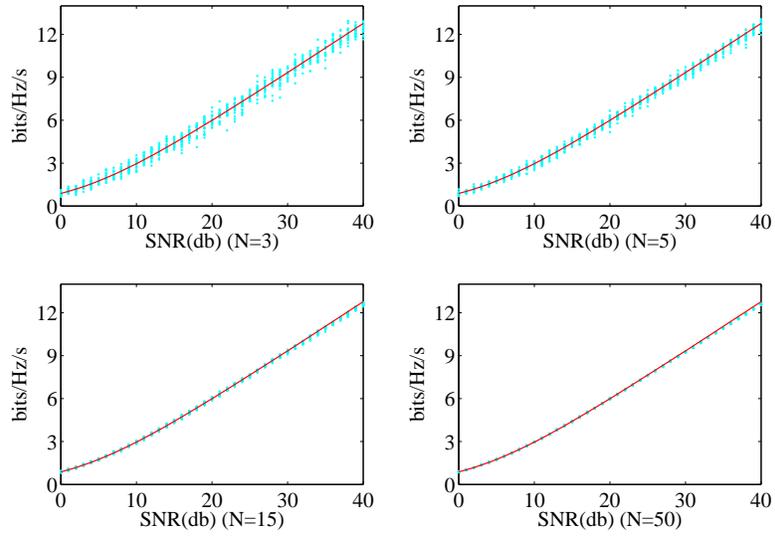}
\medskip
  \caption{\label{fig:MIMOcap}
Asymptotic capacity (solid line) of the channel with ISI, L=2, in a MIMO system with
frame length K=2 compared with the capacity of the same channel for different block sizes
(dots). }
\end{figure}

In this subsection, the asymptotic capacity of a slow-fading MIMO channel with $L=2$ and
the frame length of $K=2$ in different SNR is compared with the capacity of such a
channel for several $N$. The channel matrix for this system is as follows:
$$H=\left[\begin{matrix}
A& B &0\\
0& A& B
  \end{matrix}\right],$$
and the results are depicted in Fig. \ref{fig:MIMOcap}. As the figure shows, with
increasing the size of the blocks, the system capacity fast approaches the asymptotic
capacity, suggesting a reasonable match between the asymptotic capacity and the capacity
with a block size of 10 and bigger.

\section{Appendix I: Prerequisites}
\subsection{Notations}\label{sec:notations}
The following notations are used in the paper:
\begin{tabbing}
test, \= jhk \= lkjd \kill
$M_d(\CC)$\>\>\text{complex $d\times d$ matrices}\\
$M_d(\cA)$\>\>\text{$d\times d$ matrices with entries from the algebra $\cA$}\\
$[D]_{ij}$\>\>\text{$i,j$-entry of the matrix $D$}\\
$\tr_d$\>\>\text{normalized trace on $M_d(\CC)$}\\
$\Im\left(X\right)$\>\>$\mbox{Imaginary part of
} X$\\
$I_d$\>\>$d\times d \mbox{ Identity matrix}$\\
$\id$\>\>$\text{identity operator on a Hilbert space}$\\
$\overline{X}$\>\>$\mbox{complex conjugate of } X$\\
$\delta_{ij}$\>\>$\mbox{Dirac delta function}$\\
$X^*$\>\>$\mbox{Hermitian conjugate of matrix }X$\\
$\CC^+$\>\>$\mbox{complex upper half plane}$
\end{tabbing}
The Cauchy transform of a probability measure $\mu$ on $\RR$ is defined by
$$G(z)=\int_\RR \frac 1{z-t}d\mu(t)\qquad (z\in\CC^+).$$

\subsection{(Operator-valued) non-commutative probability spaces and freeness}
A pair $(\cA,\ff)$ consisting of a unital algebra and a linear functional $\ff:\cA\to\CC$
with $\ff(1)=1$ is called a non-commutative probability space. If $\cB$ is a subalgebra
of $\cA$, then a mapping $E:\cA\to\cB$ is called a conditional expectation if we have for
all $a\in\cA$ and $b_1,b_2\in\cB$ that
$$E[b_1ab_2]=b_1E[a]b_2.$$
An algebra $\cA$ with a conditional expectation onto a subalgebra $\cB$ is called a
$\cB$-valued probability space.

If we are given such a $\cB$-valued probability space then we say that unital subalgebras
$\cA_i\subset \cA$ ($i\in I$) are free over $\cB$ (or with respect to $E$) if the
following is satisfied: whenever we have $a_1,\dots,a_n\in\cA$ such that
$a_j\in\cA_{i(j)}$ ($i(1),\dots,i(n)\in I$) with $i(1)\not=i(2)$, $i(2)\not= i(3)$,
\dots, $i(n-1)\not=i(n)$ and with $E[a_j]=0$ for all $j=1,\dots,n$, then we also have
that $E[a_1\cdots a_n]=0$. In the case that $\cB=\CC$ (i.e., $E$ is just a linear
functional $\ff$) we say that the $\cA_i$ are free.

Elements in $\cA$ are called free (over $\cB$), if the algebras generated by them are
free (over $\cB$); they are called $*$-free (over $\cB$), if the $*$-algebras generated
by them are free (over $\cB$).

\subsection{Convergence in distribution}
Let $(\cA_N,\ff_N)$ ($N\in\NN$) and $(\cA,\ff)$ be non-commutative probability spaces.
Let $I$ be an index set and consider for each $i\in I$ random variables
$a_N^{(i)}\in\cA_N$ and $a_i\in\cA$. We say that $(a_N^{(i)})_{i\in I}$ converges in
distribution to $(a_i)_{i\in I}$ and denote this by
$$(a_N^{(i)})_{i\in I}\tovert (a_i)_{i\in I},$$ if we have that
each joint moment of $(a_N^{(i)})_{i\in I}$ converges to the corresponding joint moment
of $(a_i)_{i\in I}$, i.e.~ if we have for all $n\in\NN$ and all $i(1),\dots,i(n)\in I$
\begin{equation}
\lim_{N\to\infty}\ff_N(a_{N}^{(i(1))}\cdots a_{N}^{(i(n))})= \ff(a_{i(1)}\cdots
a_{i(n)}).
\end{equation}
We say that $(a_N^{(i)})_{i\in I}$ converges in $*$-distribution to $(a_i)_{i\in I}$ if
$(a_N^{(i)},a_N^{(i)*})_{i\in I}$ converges in distribution to $(a_i,a_i^*)_{i\in I}$.

\subsection{Operator-valued semicircular elements}\label{sec:semi}
Let $(\cA,E:\cA\to\cB)$ be a $\cB$-valued probability space and let, in addition, be
given a linear mapping $\eta:\cB\to\cB$. Then an element $s\in\cA$ is called a
$\cB$-valued operator-valued semicircular element with covariance mapping $\eta$ if one
has $E[sbs]=\eta(b)$ for all $b\in\cB$ and, more generally, for all $m\in\NN$ and all
$b_1,\dots,b_{m-1}\in\cB$ that
$$E[sb_1s\cdots sb_{m-1}s]=\sum_{\pi\in NC_2(m)}\eta_\pi[b_1,\dots,b_{m-1}],$$
where $NC_2(m)$ are the non-crossing pairings of $m$ elements (for details on
non-crossing pairings in the context of free probability see \cite{Nica-Speicher}) and
where $\eta_\pi$ is given by an iterated application of the mapping $\eta$ according to
the nesting of the blocks of $\pi$. If one identifies a non-crossing pairing with a
putting of brackets at the positions of the $s$'s, then the way that $\eta$ has to be
iterated is quite obvious. To make this clear, let us consider as an example just the
contribution of the five non-crossing pairings of six elements to the sixth moment. The
latter is given by
\begin{align*}
E[sb_1sb_2sb_3s&b_4sb_5s]=\eta(b_1)\cdot b_2 \cdot\eta(b_3)\cdot b_4\cdot \eta(b_5)\\
&+\eta(b_1)\cdot b_2\cdot\eta\bigl(b_3\cdot\eta(b_4)\cdot b_5 \bigr)+\eta\Bigl(b_1\cdot
\eta\bigl(b_2\cdot\eta(b_3)\cdot b_4\bigr)\cdot b_5\Bigr)\\&+
\eta\bigl(b_1\cdot\eta(b_2)\cdot b_3\bigr)\cdot b_4\cdot\eta(b_5)+
\eta\bigl(b_1\cdot\eta(b_2)\cdot b_3\cdot\eta(b_4)\cdot b_5\bigr),
\end{align*}
corresponding to:
$$
\begin{matrix}
\text{\usebox{\NCseZ}\qquad\qquad\qquad}&\text{\usebox{\NCszZ}}\qquad\qquad\qquad
\\
\eta(b_1)\cdot b_2 \cdot\eta(b_3)\cdot b_4\cdot \eta(b_5)& \eta(b_1)\cdot
b_2\cdot\eta\bigl(b_3\cdot\eta(b_4)\cdot b_5\bigr)
\end{matrix}
$$
\vskip1cm
$$
\begin{matrix}
\text{\usebox{\NCsdZ}}\qquad\qquad\qquad&\text{\usebox{\NCsvZ}}\qquad\qquad\qquad\\
\eta\bigl(b_1\cdot\eta(b_2)\cdot b_3\bigr)\cdot b_4\cdot\eta(b_5)& \eta\bigl(b_1\cdot
\eta(b_2)\cdot b_3\cdot\eta(b_4)\cdot b_5\bigr)
\end{matrix}
$$
\vskip1cm
$$
\begin{matrix}
\text{\usebox{\NCsfZ}}\qquad\qquad\qquad\\
\eta\Bigl(b_1\cdot\eta\bigl(b_2\cdot\eta(b_3)\cdot b_4\bigr)\cdot b_5\Bigr)
\end{matrix}
$$

For the rigorous definition of $\eta_\pi$ and more details on operator-valued
semicircular elements, we refer to \cite{Spe}.

In the situations which are relevant to us, the algebras $\cA$ and $\cB$ are operator
algebras of bounded operators on Hilbert spaces. In such a case, the main statement about
an operator-valued semicircular element $s$ is the following description of its
operator-valued Cauchy transform. Define
$$\cG:\CC^+\to\cB,\qquad \cG(z):=E[\frac 1{z-s}].$$
This is an analytic map in the upper half plane and it is, for any $z\in\CC^+$,
determined by the operator-valued quadratic equation
\begin{equation}\label{eq:oper-G}
z \cG(z)=\id+\eta\bigl(\cG(z)\bigr)\cdot\cG(z).
\end{equation}
For a derivation and details on this, see \cite{Voi,Spe}. In \cite{HFS} it is shown that Eq.
\eqref{eq:oper-G} has for fixed $z\in\CC^+$ exactly one solution $\cG$ with negative imaginary
part; furthermore, this solution is the limit of iterates $\cG_n=\cF_z^n(\cG_0)$ for any
initial point $\cG_0$ with negative imaginary part. $\cF_z$ is here the mapping
\begin{equation}\label{eq:iteration}
\cF_z(\cG)=\bigl(z\cdot\id-\eta(\cG)\bigr)^{-1}.
\end{equation}

\section{Appendix II: Proof of the main theorems}\label{sec:proof}

\subsection{Proof of Theorem \ref{thm:square}}
There are several alternative methods of proof of Theorem \ref{thm:square}. It can be
derived from Girko \cite{G} by specializing his Theorem to a block matrix with $N^2$
blocks of size $d\times d$ obtained from our matrix $X_N$ by a suitable similarity
transformation. It can be derived by elementary method of moments, see \cite{RTBS}. We
choose here to give a proof by using various results from the theory of operator-valued
free probability, thus showing how this result fits conceptually into the frame of
operator-valued free probability. The connection between random matrices and
operator-valued free probability (for ``band'' random matrices, with independent entries
but variances depending on the position of the entry) was made by Shlyakhtenko in
\cite{Shlyakhtenko-96,Shlyakhtenko-97}.

First, one has to observe that the blocks of $X_N$ converge almost surely to a semi-circular
family (see \cite{VDN-92,Hiai-Petz,Bryc-Dembo-Jiang,Hiai-Petz00a}), thus the wanted limit
distribution of $X_N$ is the same as the one of a $d\times d$-matrix $S$, where the entries of
$S$ are from a semi-circular family, with covariance $\sigma$. By using the description of
operator-valued cumulants of this matrix in terms of the cumulants of the entries of the matrix
(see \cite{Nica-Shlahtenko-Speicher02b}), it is readily seen that $S$ is a $M_d(\CC)$-valued
semi-circular element, with covariance $\eta$. The equation for $\cG(z)$ follows then from the
basic $R$-transform or cumulant theory of operator-valued free probability theory, see
Sect.~\ref{sec:semi} above, in particular Eq. \eqref{eq:oper-G}.

\subsection{Proof of Theorem \ref{thm:H-non-sep}}\label{proof:6-2}

If we decompose the positive definite matrices $\Psi^{(s)}$ and $\hat\Psi^{(s)}$ as
$\hat\Psi^{(s)}=A_s^2$ and $\Psi^{(s)}=B_s^2$ (where we take the positive square roots
$A_s=\sqrt{\hat\Psi^{(s)}}$ and $B_s=\sqrt{\Psi^{(s)}}$) then our $n\times n$ matrix $H_n$ can
be written as
$$H_n=\sum_{s=1}^t A_s Z_s B_s$$
where $Z_1,...,Z_t$ are independent $n\times n$ matrices of independent complex Gaussian
variables.

By our assumption on the convergence of $(\Psi^{(s)},\hat\Psi^{(s)})_{s=1,\dots,t}$ we know
that also $(A_s,B_s)_{s=1,\dots,t}$ converges in distribution to $(a_s,b_s)_{s=1,\dots,t}$,
where $a_s=\sqrt{\Psi_s}$ and $b_s=\sqrt{\hat\Psi_s}$. By the asymptotic freeness of Gaussian
random matrices from non-random matrices \cite{VDN-92,Nica-Speicher,Hiai-Petz,Hiai-Petz00a} we
know then that $(A_s,B_s,Z_s)_{s=1,\dots,t}$ converges in $*$-distribution to
$(a_s,b_s,c_s)_{s=1,\dots,t}$ in some $(\cA,\ff)$, where $c_1,\dots,c_t$ are $*$-free circular
elements such that $a_1,b_1,\dots,a_t,b_t$ is $*$-free from $c_1,\dots,c_t$. (A circular
element is of the form $c=s_1+is_2$ where $s_1,s_2$ are free semicircular elements.) By $\cB$
we denote, as in our theorem, the subalgebra of $\cA$ which is generated by all
$a_1,b_1,\dots,a_t,b_t$.

Then $H_n$ converges in $*$-distribution to
$$H=\sum_{s=1}^t a_sc_sb_s,$$
and $H_nH_n^*$ converges to $HH^*$. To calculate the distribution of $HH^*$ we go again
over to the selfadjoint $2\times 2$ matrix
$$X=\begin{bmatrix}
0&H\\H^*&0\end{bmatrix} =\sum_{s=1}^t \begin{bmatrix} 0&a_sc_sb_s\\b_sc_s^*a_s&0
\end{bmatrix}=\sum_{s=1}^t\begin{bmatrix} a_s&0\\0&b_s\end{bmatrix}
\begin{bmatrix} 0&c_s\\c_s^*&0\end{bmatrix}
\begin{bmatrix} a_s &0\\0&b_s\end{bmatrix}.$$
The relation between the distribution of $X$ and the distribution of $HH^*$ is as in
Section \ref{proof:H-block}, thus it remains essentially to determine the distribution of
$X$.

Put
$$A_s:=\begin{bmatrix} a_s&0\\0&b_s\end{bmatrix}$$
and
$$S_s:=\begin{bmatrix} 0&c_s\\c_s^*&0\end{bmatrix},$$
so that we have
$$X=\sum_{s=1}^t A_s S_s A_s$$

The main problem is now that the different terms $A_sS_sA_s$ in $X$ are not free and thus one
cannot reduce the situation directly to the case $t=1$. However, we have operator-valued
freeness with respect to a suitably chosen conditional expectation. Namely, let us first take
the conditional expectation $E$ from $\cA$ to $\cB$ (which exists by general arguments, because
we are in a tracial situation, see, e.g., \cite{Voi}) and then we go over to $2\times 2$
matrices by taking this $E$ entrywise, i.e. we consider
$$1\otimes E:M_2(\cA)\to M_2(\cB)$$
given by
$$1\otimes E \begin{bmatrix} a_1&a_2\\a_3&a_4\end{bmatrix}=
\begin{bmatrix} E(a_1)&E(a_2)\\E(a_3)&E(a_4)\end{bmatrix}.$$

From Theorem 3.5 in \cite{Nica-Shlahtenko-Speicher02b} it follows now that, for each
$s=1,\dots,t$, $A_sS_sA_s$ is a semicircular element over $M_2(\cB)$, and furthermore, that all
$A_1S_1A_1,\dots,A_tS_tA_t$ are free over $M_2(\cB)$. But this implies that also their sum $X$
is a semicircular element over $M_2(\cB)$. It remains to calculate its covariance function. We
have
$$\eta(D)=1\otimes E (XDX),$$
i.e., for $d_1,d_2,d_3,d_4\in\cB$
$$
\eta\begin{bmatrix}d_1&d_3\\d_4&d_2\end{bmatrix}=\sum_{i,j=1}^t
\begin{bmatrix}
E[a_ic_ib_id_2b_jc_j^*a_j]& E[a_ic_ib_id_4a_jc_jb_j]\\
E[b_ic_i^*a_id_3b_jc_j^*a_j]&E[b_ic_i^*a_id_1a_jc_jb_j]
\end{bmatrix}
$$
It is quite easy to see that the conditional expectation $E:\cA\to\cB$ acts for all
$b\in\cB$ as
$$E[c_ibc_j^*]=\delta_{ij}\ff(b)$$
$$E[c_i^*bc_j]=\delta_{ij}\ff(b)$$
$$E[c_i^*bc_j^*]=0$$
$$E[c_ibc_j]=0$$
Thus
$$\eta\begin{bmatrix}d_1&d_3\\d_4&d_2\end{bmatrix}=
\begin{bmatrix}
\sum_{i=1}^t a_ia_i\ff(b_id_2b_i)&0\\
0& \sum_{i=1}^t b_ib_i \ff(a_id_1a_i)\end{bmatrix}= \begin{bmatrix}
\eta_2(d_2)&0\\0&\eta_1(d_1)
\end{bmatrix}$$
where $\cB_1$ is the algebra generated by $a_1^2,\dots,a_t^2$ (i.e., the algebra generated by
$\Psi_1,\dots,\Psi_t$) and $\cB_2$ is the algebra generated by $b_1^2,\dots,b_t^2$ (i.e, the
algebra generated by $\hat\Psi_1,\dots,\hat\Psi_t$) and $\eta_1:\cB_1\to\cB_2$ and
$\eta_2:\cB_2\to\cB_1$ are the mappings as in our Theorem.

Denote by $\cD$ the subalgebra of $M_2(\cB)$ of the form
$$\cD=\begin{bmatrix} \cB_1&0\\0&\cB_2
\end{bmatrix}.$$
We see that $\eta$ maps $\cD$ to itself. Then it follows by Theorem 3.1 of
\cite{Nica-Shlahtenko-Speicher02b} that $X$ is also a semicircular element over $\cD$,
with the same $\eta$. This implies then that our corresponding operator-valued Cauchy
transform $\cG(z)$ lies in $\cD$, thus is of the form
$$\cG(z)=\begin{bmatrix}
\cG_1(z)&0\\
0&\cG_2(z)\end{bmatrix},$$ where $\cG_1(z)\in\cB_1$ and $\cG_2(z)\in\cB_2$. The rest
follows then from analyzing the corresponding operator-valued quadratic equation
\eqref{eq:oper-G} for $\cG(z)$ and relating it with the Cauchy transform of $HH^*$ as in
Section \ref{proof:H-block}.

\section{Appendix III: Selfadjoint case with rectangular blocks}
In some applications one encounters situations where the blocks themselves might not be
square matrices, but more general rectangular matrices. Of course, the sizes of the
blocks must fit together to make up a big square matrix. This means that in Theorem
\ref{thm:H-block} we replace $n=dN$ by a decomposition $n=N_1+\cdots +N_d$, and the block
$A^{(i,j)}$ will then be a $N_i\times N_j$-matrix. We are interested in the limit that
$N_i/n$ converges to some number $\alpha_i$.

Let us first introduce the generalizations of our relevant notations from the square
case. Note that dependent rectangular blocks can be re-cut into different nonequivalent
configurations of dependent blocks. We will assume that such repartitioning has already
been done and resulted in
 the covariance function $\sigma(i,j;k,l)$ that can only be different from zero if
the size of the block $A^{(i,j)}$ fits (at least in the limit $n\to\infty$) with the size
of the block $A^{(k,l)}$.

\begin{notation}
Fix a natural number $d$ and a $d$-tuple $\alpha=(\alpha_1,\dots,\alpha_d)$ with
$0<\alpha_i<1$ for all $i=1,\dots,d$ and $\alpha_1+\cdots+\alpha_d=1$. Furthermore, let a
covariance function $\sigma=\bigl(\sigma(i,j;k,l)\bigr)_{i,j,k,l=1}^d$ be given with the
property that \eqref{symmetry} holds and in addition $\sigma(i,j;k,l)=0$ unless
$\alpha_i=\alpha_l$ and $\alpha_j=\alpha_k$. Then we use the following notations.

1) $M_\alpha(\CC)$ are those matrices from $M_d(\CC)$ which correspond to square blocks,
$$M_\alpha(\CC):=\{D\in M_d(\CC)\mid
\text{$[D]_{ij}=0$ unless $\alpha_i=\alpha_j$}\}.$$

2) We define the \emph{weighted covariance mapping}
$$\eta_\alpha:M_\alpha(\CC)\to M_\alpha(\CC)$$ as follows:
$$\left[\eta_\alpha(D)\right]_{ij}:=\sum_{k,l=1}^d \sigma(i,k;l,j)\cdot\alpha_{k}\cdot [D]_{kl}.$$

3) Furthermore, the weighted trace
$$\tr_\alpha: M_\alpha(\CC)\to \CC$$
is given by
$$\tr_\alpha(D):=\sum_{i=1}^d \alpha_i \cdot[D]_{ii}.$$

\end{notation}

The following statement for rectangular blocks can be reduced to the case of square
blocks by cutting the rectangular blocks into smaller square blocks (at least
asymptotically); for a more direct combinatorial proof, see \cite{RTBS}.

\begin{theorem}\label{thm:rectangular}
With the above notation, for $\left\{N_1,\dots,N_d\right\}\subset\NN$ consider block
matrices
$$X_{N_1,\dots,N_d}=\bigl(A^{(i,j)}\bigr)_{i,j=1}^d.$$
For each $i,j=1,\dots,d$, the $A^{(i,j)}$ are Gaussian $N_i\times N_j$ random matrices,
$A^{(i,j)}=\bigl(a^{(i,j)}_{rp}\bigr)_{r=1,\dots,N_i\atop p=1,\dots,N_j}$. The latter are
such that the collection of all entries $\{a_{rp}^{(i,j)}\mid i,j=1,\dots,d,\,
r=1,\dots,N_i, p=1,\dots,N_j\}$ of the matrix $X_{N_1,\dots,N_d}$ forms a Gaussian family
which is determined by
$$a^{(i,j)}_{rp}=\overline{a_{pr}^{(j,i)}}\qquad\text{for all $i,j=1,\dots,d$, $r=1,\dots,N_i$,
$p=1,\dots,N_j$}$$ and the prescription of mean zero and covariance
$$E[a_{rp}^{(i,j)} a_{qs}^{(k,l)}]=\frac 1n \delta_{rs}\delta_{pq}\cdot
\sigma(i,j;k,l),$$ where we put
$$n:=N_1+\cdots N_d.$$
Then, for $n\to\infty$ such that
$$\lim_{n\to\infty} \frac {N_i}n=\alpha_i\qquad\text{for all $i=1,\dots,d$},$$
the matrix $X_{N_1,\dots,N_d}$ has almost surely a limiting eigenvalue distribution whose
Cauchy transform $G(z)$ is determined by
$$G(z)=\tr_\alpha(\cG(z)),$$
where $\cG(z)$ is an $M_\alpha(\CC)$-valued analytic function on the upper complex half
plane, which is uniquely determined by the facts that \eqref{eq:cG lim} holds and that it
satisfies for all $z$ in the upper complex half plane the matrix equation
\begin{equation} \label{eq:rectangular}
z \cG(z)= I_d + \eta_\alpha(\cG(z))\cdot \cG(z).
\end{equation}
\end{theorem}

\section*{Acknowledgement} We thank Ralf M\"uller for bringing the problem of block
matrices in the context of MIMO to our attention. We also acknowledge the very
constructive feedback and suggestions of two reviewers of an earlier version of this
paper.

\clearpage

\clearpage

\listoffigures
\end{document}